\documentclass[supplement]{ptptex}

\usepackage{graphicx}

\title{
Structural Transitions and Magnetic Structure in NH$_{4}$CuCl$_{3}$ via $^{14}$N-NMR}

\author{
Katsuaki Kodama$^{1}$, Masashi Takigawa$^{1}$, Hidekazu Tanaka$^{2}$%
}

\inst{
$^{1}$Institute for Solid State Physics, University of Tokyo, Kashiwanoha, Kashiwa, Chiba 277-8581, Japan \\
$^{2}$Research Center for Low Temperature Physics, Tokyo Institute of Technology, Oh-okayama, Meguro-ku, Tokyo 152-8551, Japan
}

\abst{
We report results of $^{14}$N-NMR experiments on NH$_{4}$CuCl$_{3}$ 
at the magnetic field of  7~T, where the 1/4-magnetization plateau is observed at
low temperatures.  The quadrupole splitting parameter $\nu_{z}$ splits 
below  70~K, indicating a structural phase transition.  At 4.2~K, eight 
N sites with distinct values of both $\nu_{z}$ and the magnetic hyperfine shift
$K_{z}$ are resolved in the NMR spectrum for general field directions.  We then
conlude that the magnetic structure in the 1/4-plateau does not break the 
symmetry of the crystal.  Based on the NMR and the recent neutron scattering 
results by R\"{u}egg \textit{et al}. [Phys. Rev. Lett. 93 (2004) 037207], we propose
that triplet dimers in the 1/4-plateau is formed not between the nearest neighbor pairs
but over different chains.}

\begin{document}

\maketitle

Magnetization plateaus in quantum spin systems have attracted strong recent interest 
as a novel example of quantum many body effects. Oshikawa formulated a 
necessary condition for magnetization plateaus in arbitrary
dimensions~\cite{oshikawa}, 
which in certain cases predicts that the periodicity of the 
ground state wave function is larger than the periodicity of the crystal.  
Such a magnetic superlattice should be a consequence of localization 
of magnetic excitations due to repulsive interactions and bear common physics to a
number of interesting quantum many body phenomena such as Mott transition and charge
ordering.  Spontaneous symmetry breaking was indeed observed in the 
1/8 magnetization plateau phase of the frustrated 2D dimer system 
SrCu$_{2}$(BO$_{3}$)$_{2}$~\cite{kodama}.  NH$_{4}$CuCl$_{3}$ 
has been considered as another candidate since magnetization plateaus were observed 
at 1/4 and 3/4 of the saturated magnetization but not at 1/2~\cite{shira}.   

\begin{figure}[t]
\centering
\includegraphics[scale=0.5]{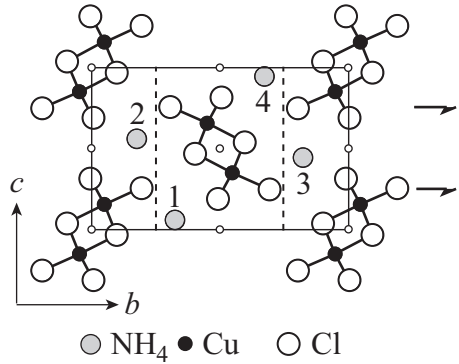}
\caption{Crystal structure of NH$_{4}$CuCl$_{3}$ at room temperature (space group 
P2$_{1}$/c) viewed along the $a$-axis.  The inversion center, the $b$-screw axis,
and the $c$-glide plane, are shown by the circles, the arrows, and the dashed lines, respectively.}
\label{fig1}
\end{figure}
The crystal structure of NH$_{4}$CuCl$_{3}$ has the space group 
P2$_{1}$/c at room temperature, in which Cu$^{2+}$ ions form zigzag chains
along the $a$-axis (Fig. 1)~\cite{structure}.  This is identical to the structure
of TlCuCl$_{3}$ and KCuCl$_{3}$, which have dimer singlet ground states
at zero magnetic field with a finite energy gap to the triplet excitations.  
The gap can be reduced by the field and an antiferromagnetic order appears 
above a critical field at which the gap vanishes.  This phenomenon is best described 
as the Bose condensation of triplets ~\cite{nikuni}.  In contrast,
NH$_{4}$CuCl$_{3}$ shows an antiferromagnetic order at
zero-field~\cite{kurniawan,ruegg} and the magnetization increases continuously from
zero field~\cite{shira}.  Since all Cu sites 
are equivalent in the P2$_{1}$/c structure, the 1/4 and the 3/4 plateaus in 
NH$_{4}$CuCl$_{3}$ must be accompanied by magnetic superstructure 
breaking the crystal symmetry if the crystal structure remains unchanged down to low
temperatures.    

In contradiction to this expectation, various experiments indicate existence of
inequivalent Cu sites with different magnetic characters.  Specific heat shows a
Schottkey-type peak at zero-field indicating that the majority of spins are in singlet
states, in addition to an anomaly at $T_{N}$=1.3~K due to antiferromagnetic ordering
of the rest of spins~\cite{kurniawan}.  Existence of singlet dimers in the ground
state was confirmed by the observation of two dispersive finite energy
excitations at 1.6~meV and 3~meV by ESR~\cite{nojiri} and neutron
scattering~\cite{oosawa} experiments. Both modes are only weakly
dispersive~\cite{oosawa}.  NMR experiments on a single crystal enriched with
$^{15}$N isotope (spin 1/2) revealed splitting of the spectrum below 70 K for the
external fields of 3 T and 6 T~\cite{shima}.  This splitting cannot be due to
formation of magnetic superstructure because the magnetization plateau does not
appear at 3~T.  Since the magnetic susceptibility shows no anomaly around 70~K, the
splitting of $^{15}$N-NMR lines is likely to be due to lowering of crystal symmetry
caused by ordering of NH$_{4}$ molecules as detected by the infrared absorption
measurements~\cite{IR}.  The structural transition at 70~K was also suggested by
elastic anomalies~\cite{sound}.  Based on these experimental results, 
Matsumoto proposed a model consisting of three distinct dimer sublattices with
different energy gaps generated by structural distortion~\cite{matsu}.  In this
model, the plateaus correspond to successive saturation of each sublattice dimers.
The recent neutron diffraction experiments~\cite{ruegg} revealed that the structural
transitions occur in two steps.  First, the intensity of (001) reflection 
begins to increases below 156 K, indicating breaking of the c-glide symmetry.  Then
the reflection at $(h k l)$ with half integer $k$ appears below 70 K, pointing to
doubling of the unit cell along the $b$-axis.  

In this paper, we report results of NMR experiments on $^{14}$N nuclei (spin 1)
in a twin free single crystal of NH$_{4}$CuCl$_{3}$ at 7 T where the 1/4 plateau 
appears at low temperatures.  The NMR spectra were obtained from the Fourier
transform of the spin-echo signal.  The $^{14}$N-NMR spectra generally consist of 
two resonance lines at the following frequencies split by the electric quadrupole 
interaction~\cite{abragam},  
\begin{equation}
f_{\pm} = (1+K_{z}) \gamma_{n} H_{0} \pm \nu_{z},  \; \nu_{z}=(3eQ/4h)V_{zz}. 
\label{NMR}
\end{equation}
Here, $\gamma_{n}$ is the nuclear gyromagnetic ratio, $Q$ is the nuclear quadrupole 
moment, $H_{0}$ is the external field, $K$ is the magnetic hyperfine shift, 
and $V_{zz}=\partial^{2}V/\partial z^{2}$ is the $zz$-component of the electric 
field gradient tensor, where $z$ is the field direction.  The values of $K_{z}$ and 
$\nu_{z}$ were determined as a function of temperature and field direction.  
While the electric field gradient reflects the local symmetry of the crystal
structure, the magnetic hyperfine shift is determined by the spin density
distribution.  $^{14}$N-NMR thus allows us to examine correlation between 
structural and magnetic properties, in contrast to $^{15}$N-NMR,
from which one can obtain only the magnetic hyperfine shift.     

The four N atoms in a unit cell are equivalent at room temperature  
because they transform each other by the symmetry operations of P2$_{1}$/c group 
(Fig. 1).  A pair of N sites should yield identical NMR spectrum if the field direction
is invariant under the symmetry operation which transforms one N site to the other.  
This is always satisfied for the pair of sites related by inversion (1 and 4, 
or 2 and 3 in Fig. 1), while those pairs related by the $b$-screw (1 and 3, or 
2 and 4) or the $c$-glide (1 and 2, or 3 and 4) give the same values of $K_{z}$
and $\nu_{z}$ when the field is parallel to the $b$-axis or in the $ac$-plane.  
The observed NMR spectra are compatible with these conditions above 160~K.  
This enabled precise alignment of the $b$-axis along the field.    
  
\begin{figure}[b]
\centering
\includegraphics[scale=0.8]{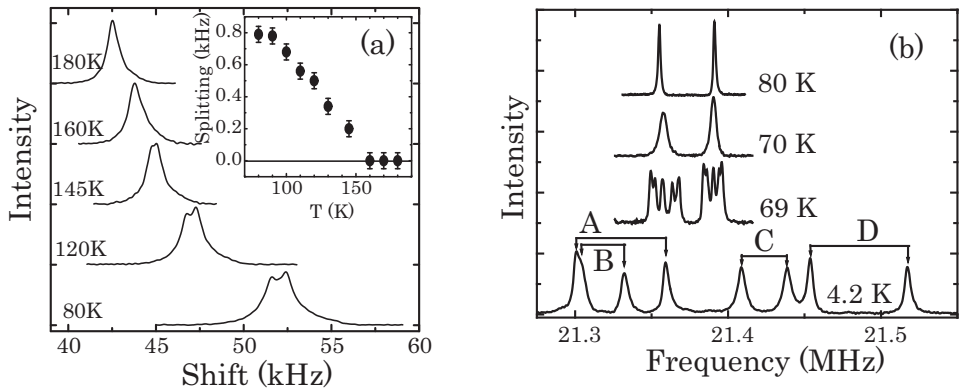} 
\caption{(a) Temperature variation of the NMR spectrum in the magnetic field of 7~T
along the $b$-axis.  Although one peak of the quadreupole split pair of lines is shown
here, the other peak also shows similar splitting below 160~K.  The interval of the
split peaks is plotted against temperature in the inset.  (b) Temperature variation
of the NMR spectrum below 80~K in the magnetic field of 7~T approximately parallel
to the $c^{*}$-axis.}
\label{fig2}
\end{figure}
With the field set along the $b$-axis and with decreasing temperature, we found that
$\nu_{b}$ begins to split below 160~K (Fig. 2 (a)).  The splitting is very small,
about 0.8~kHz at 80~K compared to $\nu_{b}$=29~kHz.  Moreover, the spectra 
show at most two set of quadrupole split lines for any field directions.  These
results led us to conclude that the $b$-screw and the $c$-glide symmetries are
marginally broken below 160~K but the inversion symmetry is preserved.  
Therefore, the space group should be P$\bar{1}$.  This is consistent with the
observation of (001) reflection by neutron diffraction~\cite{ruegg}. 

With further decreasing temperature, the NMR spectrum develops a drastic change.
We show in Fig. 2 (b) the temperature variation of the NMR spectrum for the field 
approximately parallel (within 2$^{\circ}$) to the $c*$-axis.  The spectrum at 
80~K shows only one set of lines, since the splitting of $\nu_{z}$ is accidentally
nearly zero and the splitting of $K_{z}$ is also smaller than the experimental
resolution. The lines begin to split at $T_{c}$=69~K, clearly indicating a phase
transition.  At 4.2~K, the NMR spectrum consists of eight lines.  The spin-echo
signal for these lines oscillates as a function of the separation time $\tau$ between
the $\pi$/2- and the $\pi$-pulses.  It is known that the quadrupole interaction
causes such oscillation and its frequency is equal to 2$\nu_{z}$~\cite{abe}.
This allowed us to classify the eight lines into four pairs of 
quadrupole split lines (A$\sim$D in Fig. 2) originating from the same sites.   
The values of $K_{c*}$ and $\nu_{c*}$ for each site were then determined 
from Eq.~(\ref{NMR}) at various temperatures as shown in Fig. 3.    

\begin{figure}[t]
\centering
\includegraphics[scale=0.85]{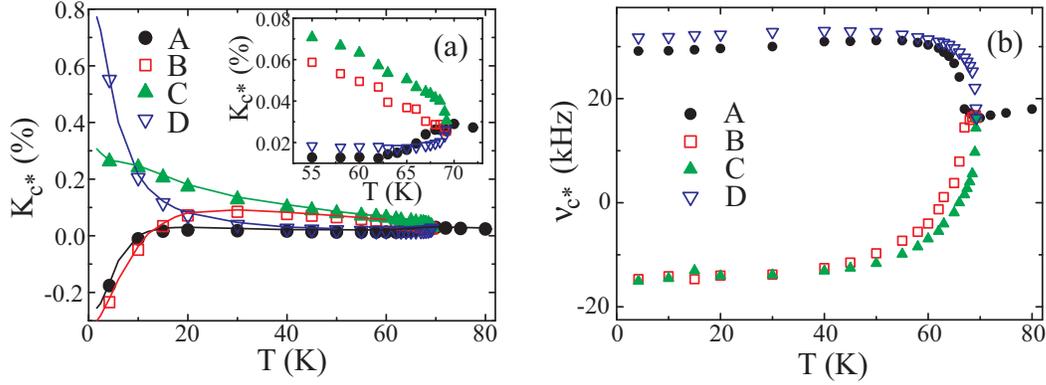}
\caption{Temperature dependence of (a) $K_{c*}$ and (b) $\nu_{c*}$ for the 
sites A$\sim$D shown in Fig. 2 in the field of 7 T approximately parallel to the
$c^{*}$-axis.}
\label{fig3}
\end{figure}
\begin{figure}[b]
\centering
\includegraphics[scale=0.7]{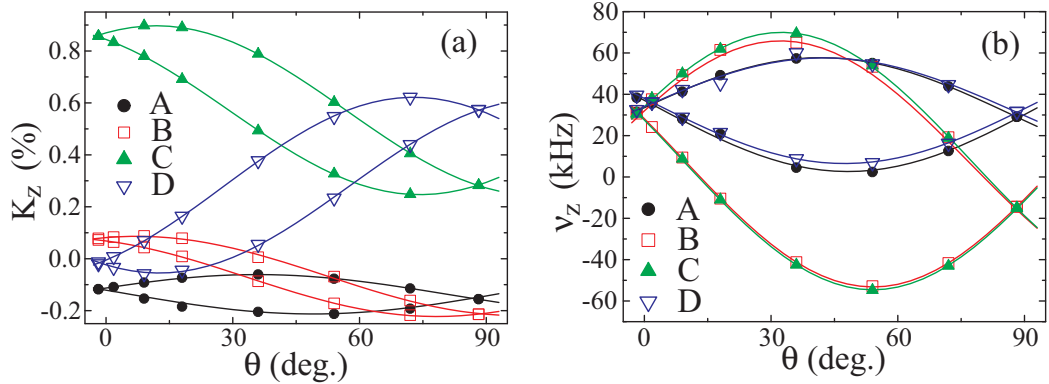}
\caption{Angular dependence of (a) $K_{z}$ and (b) $\nu_{z}$ for sites A$\sim$D in
the magnetic field of 7 T rotated in $b-c^{*}$ plane.  $\theta$ is an angle between
the magnetic field and the $b$ axis.  The lines show the fitting described in the
text.}
\label{fig4}
\end{figure}
The temperature dependences of $K_{c*}$ and $\nu_{c*}$ are strikingly different,
although both quantities begin to split at $T_{c}$.  $K_{c*}$ shows only minor
variation near $T_{c}$ but changes substantially at low temperatures.  Similar
temperature dependence was observed for $K_{b}$ by Shimaoka 
\textit{et al.}~\cite{shima}.  In contrast, $\nu_{c*}$ changes rapidly in a
temperature range close to $T_{c}$ but becomes almost $T$-independent at low
temperatures.  These results indicate a second order structural phase transition at
$T_{c}$, consistent with earlier experiments, and irrelevance of magnetic
interactions in driving the transition.  The structural change, most likely 
associated with long range orientation order of NH$_4$, would significantly modify
the exchange interactions and lead to distinct magnetic characters for different 
sites at low temperatures. We should note, however, that an isolated NH$_4$
molecule with tetrahedral symmetry does not contribute to the electric field gradient
at the N nuclei.  Close examination of the data in Fig. 3(b) reveals a peculiar
feature.  While $\nu_{Q}$ at the sites C and D changes rapidly immediately below
$T_{c}$, onset of the rapid change in $\nu_{Q}$ at the site A and B is shifted to
a lower temperature by about 2~K.  We do not have any explanation yet for such
behavior.     

The angular dependences of $K_{z}$ and $\nu_{z}$ at 4.2~K were measured with the
magnetic field rotated approximately from the $b$-axis ($\theta=0^\circ$) to the 
$c^{*}$-axis ($\theta=90^\circ$) as shown in Fig. 4.  Within the experimental
resolution, four distinct values of $K_{z}$ and $\nu_{z}$ are obtained for 
$\theta\sim0^\circ$ and $\theta\sim90^\circ$.  
Each of these values splits into two for other field directions.  The angular
dependence of $K_{z}$ and $\nu_{z}$ can be fit well to the form
$\alpha+\beta\cos(2\theta+\gamma)$, where $\alpha$, $\beta$, and $\gamma$ 
are constants, as shown by the lines in Fig. 4.  We found that the
$\theta$-dependences of both $K_{z}$ and $\nu_{z}$ for the two branches of each pair 
(A$\sim$D) are approximately symmetric with respect to
$\theta\sim0^\circ$ and $\theta\sim90^\circ$, indicating that the two branches are
approximately related by the $c$-glide operation. Therefore, breaking of 
the $c$-glide symmetry must be still marginal in the low temperature phase.  
The data of $K_{z}$ and $\nu_{z}$ both show that there are at least eight N atoms
in a unit cell.  An important consequence is that the magnetic structure
(spin density distribution) does not necessarily break the crystal symmetry.  This
is consistent with the various experimental evidences mentioned earlier that the 
1/4-plateau state emerges at low temperatures without symmetry breaking.

Now we discuss the magnetic structure of the 1/4-plateau based on the present NMR
results and the neutron results by R\"{u}egg \textit{et al}.~\cite{ruegg}.
According to the neutron data, unit cell doubling occurs along the $b$-direction but
not along other directions.  Then the unit cell of the low temperature phase contains
eight N atoms. If these sites were paired by inversion symmetry, only four sites should
be distinguished by NMR for any field direction.  From our observation of 
eight distinct values of $\nu_{Q}$ and $K_{z}$ for general field directions, we  
conclude that the inversion symmetry must be broken in the low temperature phase, 
i.e. the space group should be P1. Although R\"{u}egg \textit{et al}. concluded 
P$\bar{1}$ symmetry~\cite{ruegg}, their argument was based 
on the earlier $^{15}$N-NMR data~\cite{shima}, which reported spectra only 
along the $b$-axis.  

\begin{figure}[tbh]
\centering
\includegraphics[scale=0.7]{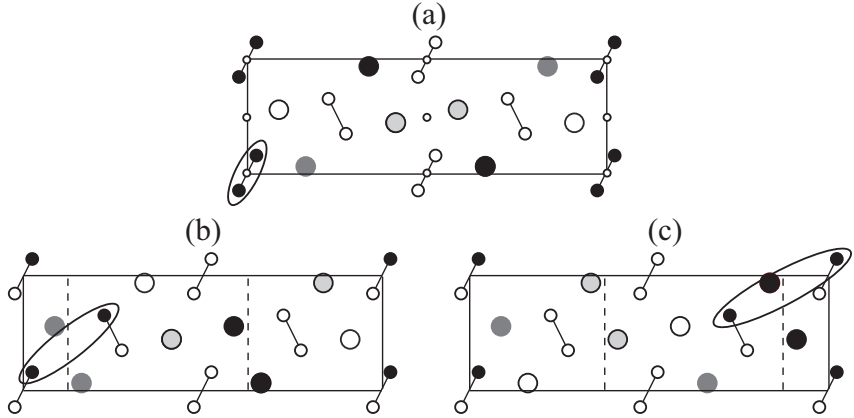}
\caption{Possible magnetic structure of the 1/4 plateau state. The small filled
(open) circles show the triplet (singlet) Cu sites.  Large circles show the N atoms.
The dashed lines in (b) and (c) indicate the approximate $c$-glide plane preserved
in each spin configuration.}
\label{fig5}
\end{figure}
We assume that in the 1/4-plateau state, 1/4 of the Cu spins form nearly fully
polarized triplet while the rest of spins are in singlet states, as suggested by
various experiments mentioned earlier.  The question then is where the fully
polarized spins are located.  All previous studies~\cite{matsu, ruegg}
assumed that triplets are formed over 1/4 of the nearest neighbor dimers as shown
in Fig. 5(a).  This spin configuration, however, preserves inversion symmetry but 
completely destroys the $c$-glide symmetry.  Since the magnetic hyperfine field at
the N sites should be determined primarily by the arrangement of fully polarized
spins, this model is clearly incompatible with the approximate $c$-glide
symmetry evidenced by the angular dependence of $K_{z}$.  This configuration also
yields similar values of $K_{z}$ for those pairs related by inversion, leading to 
only four distinct values of $K_{z}$.  This is again inconsistent with the
observation of eight distinct values of $K_{z}$ for general field directions.  

This lead us to propose that the triplet dimers must be formed over different chains
as shown in Fig. 5 (b) or (c).  The spin configuration in Fig. 5 (b) (Fig. 5(c))
has the $c$-glide plane at $y$=1/8 and 5/8 ($y$=3/8 and 7/8).  Therefore, the angular
dependence of $K_{z}$ should also satisfy the approximate glide symmetry, consistent
with our observation.  Also the inversion symmetry is completely destroyed in these 
configurations, accounting for eight distinct values of $K_{z}$ for general field
directions.  Although  the four different symbols for NH$_{4}$ molecules in 
Fig. 5 (b) and (c) should correspond to the sites A$\sim$D identified by our NMR data 
(Figs. 3 and 4), unambiguous assignment has not been made yet. 
At first sight, triplet dimers over such a long distance might appear very unlikely,
since the strongest exchange interaction in TlCuCl$_{3}$ and KCuCl$_{3}$ is located 
between nearest neighbor pairs, stabilizing the dimer singlet
state~\cite{oosawa2,kcucl,kcucl2}.  
However, we note that NH$_{4}$CuCl$_{3}$ and other two compounds may have
completely different exchange coupling scheme, because there is no similarity
at all in the magnon dispersion observed by neutron scattering experiments.
Moreover, since NH$_{4}$ molecules are located near the exchange path stabilizing
the triplet dimers in Fig. 5 (b) and (c), orientation of NH$_{4}$ molecules should 
significantly modify the exchange coupling, accounting for the extremely strong
spin-lattice coupling in this compound.       

In conclusion, our $^{14}$N-NMR experiments revealed that the structural transition
at 69~K produces sufficiently large number of inequivalent Cu sites at low
temperatures so that 1/4-magnetization plateau can emerge without symmetry breaking.
The angular dependences of the magnetic hyperfine shift and the 
quadrupole splitting parameters combined with the published neutron results 
strongly suggest that the fully polarized triplets in the 1/4-plateau are formed not 
between the nearest neighbor pairs but over distant Cu spins on different chains.

We would like to thank Yoshiyuki Shimaoka, Takao Goto, Takayuki Goto, and 
Kazuhisa Kakurai for stimulating discussions.
This work was supported by a Grant-in-Aid for Scientific Research in Priority Area
"Field-Induced New Quantum Phenomena in Magnetic Systems" from the MEXT Japan. .


%


\begin{thebibliography}{99}
  
\bibitem{oshikawa}
M. Oshikawa, M. Yamanaka, and I. Affleck, \PRL{84,2002,054429}.
\bibitem{kodama}
K. Kodama, M. Takigawa, M. Horvatic, C. Berthier, H. Kageyama, Y. Ueda, S. Miyahara, F. Becca, and F. Mila, 
\JL{Science,298,2002,395}.
\bibitem{shira}
W. Shiramura, K. Takatsu, B. Kurniawan, H. Tanaka, H. Uekusa, Y. Ohashi, K. Takizawa,
H. Mitamura, and T. Goto, \JPSJ{67,1998,1548}.
\bibitem{structure}
R. D. Willett, C. Dwiggins, R. F. Kruh, and R. E. Rundle, J. Chem. Phys. \textbf{38} (1963), 2429. 
\bibitem{nikuni}
T. Nikuni, M. Oshikawa, A. Oosawa, and H. Tanaka, \PRL{84,2000,5868}.
\bibitem{kurniawan}
B. Kurniawan, M. Ishikawa, T. Kato, H. Tanaka, K. Takizawa, and T. Goto, J. Phys. :Condens. Matter \textbf{11} (1999), 9073.
\bibitem{ruegg}
Ch. R\"{u}egg, M. Oettli, J. Schefer, O. Zaharko, A. Furrer, H. Tanaka, K. W. Kramer, H.-U. Gudel, P. Vorderwisch, 
K. Habicht, T. Polinski, and M. Meissner, \PRL{93,2004,037207}.
\bibitem{nojiri}
H. Nojiri, H. Tanaka, B. Kurniawan, and M. Motokawa, in 
{\it Proceeding of French-Japanese Symposiumu on Qunatum Properties od Low-Dimensional Antiferromagnets},
edited by Y. Ajiro and J. P. Boucher (Kyushuu University Press, Fukuoka, 2002).
\bibitem{oosawa}
A. Oosawa, T. Ono, K. Kakurai, and H. Tanaka, in 
{\it Proceeding of the Workshop on the Perspectives in Single Crystal Neutron Spectroscopy}, 
(ILL Publications, Grenoble, 2003), cond-mat/0304172.
\bibitem{shima}
Y. Shimaoka, T. Goto, K. Kodama, M. Takigawa, and H. Tanaka, Physica B\textbf{329-333} (2003), 894.
\bibitem{IR}
A. M. Heyns and C. J. H. Schutte, J. Mol. Struct. {\bf 8}, (1971), 339.
\bibitem{sound}
S. Schmidt, S. Zherlitsyn, B. Wolf, H. Schwenk, B. Luthi, and H. Tanaka, Europhys. Lett. \textbf{53} (2001), 591.
\bibitem{matsu}
M. Matsumoto, Phys. Rev. B{\bf 68} (2003), 180403.
\bibitem{abragam}
A. Abragam, \textit{The Principle of Nuclear Magnetism}, (Oxford Clarendon Press, 1961).
\bibitem{abe}
H. Abe, H. Yasuika, M. Matsuura, A. Hirai, and T. Shinjo, \JPSJ{19,1964,1491}.
\bibitem{oosawa2}
A. Oosawa, T. Kato, H. Tanaka, K. Kakurai, M. M\"{u}ller, and H.-J. Mikeska, \PRB{65,2002, 094426}. 
\bibitem{kcucl} 
N. Cavadini, W. Henggler, A. Furrer, H.-U. G\"{u}del, K. Kramer, and H. Mutka, Eur. Phys. J. B\textbf{7} (1999), 519.  
\bibitem{kcucl2}
T. Kato, A. Oosawa, H. Tanaka, K. Nakajima, and K. Kakurai, J. Phys. Soc. Jpn. Suppl. A\textbf{70} (2001), 160.


\end{thebibliography}
\end{document}